\documentclass{article}
\usepackage[ansinew]{inputenc}
\usepackage{amsmath,amssymb,amsthm,amscd}
\usepackage{hyperref}
\usepackage{here, caption}
\usepackage{graphicx}

\newtheorem{definition}{Definition}[section]

\begin{document}

\title{Properties of the thermal two-point functions in curved spacetimes for a self-interacting scalar field}
\author{Samuel Rutili\\
Dipartimento di Fisica\\ Universit\`a degli Studi di Pavia \& INFN - Sezione di Pavia\\ Via Bassi, 6, I-27100 Pavia, Italy.\\
\texttt{samuel.rutili@pv.infn.it}
}
\maketitle

\begin{abstract}
  QFT is one of the most succesful theories in physics. It allows to reach very precise predictions concerning physical systems in a relativistic regime, on flat spacetime.
  If the spacetime is curved the traditional approach to QFT is no longer possible, since the lack of a symmetry group (replacing the Poincar\'e group of the flat case) leads to the lack of a preferred Hilbert space as a founding object of the theory (the analogous of the Fock space in QFT). So a new point of view is needed: The quantum fields are no longer interpreted as operators on a Hilbert space, but as abstract objects defined only by some physical requirements. The physical observables are combinations of such fields. The natural mathematical framework to formalize these ideas is the so called Algebraic Quantum Field Theory (AQFT). In such a mathematical context it is possible to describe thermal (KMS) quantum states in a consistent way. We are interested in a particular case of curved spacetime: The spacetime generated by a black hole (Schwarzschild spacetime). We will focus our attention on a particular field: The interacting massive scalar field.
  A crucial point to describe a quantum field on Schwarzschild spacetime is to study the convergence of the two-point function, which is the fundamental object one needs to compute the espectation values of the observables of the theory. This is the main goal of this work (joint work with C. Dappiaggi).
\end{abstract}


\section{Flat spacetime} \label{flat}

The first step of our discussion is a brief review of K. Fredenhagen's and F. Lindner's work (see \cite{FalkThesis}). The aim of this section is to point out the main properties of the KMS states on Minkowski spacetime for a massive scalar field, in order to generalize them to the curved case later.
To define a KMS state we need first to define a state in the algebraic framework. For a complete overview of the algebraic approach, see \cite{book}. Here we just remind that:
\begin{definition}
A free massive scalar field on a manifold $(M,g)$ is a function $\phi :M\rightarrow \mathbb{R}$, such that $P\phi (x)=0$ with $P=\Box _g-m^2$ Klein-Gordon operator, $\Box _g=\nabla _\mu \nabla ^\mu$ and $g$ metric on $M$.
\end{definition}
\begin{definition}
The algebra of observables on the manifold $(M,g)$ is the $*$-algebra $\mathcal{A}(M)$ generated by smeared fields $\phi (f):=\int \phi(x)f(x)d\mu _g $ ($f\in \mathcal{D}(M)$), encoding locality, causality and the canonical commutation relations (for details see \cite{FalkThesis}).
\end{definition}
\begin{definition}
An algebraic state is a functional $\omega :\mathcal{A}\rightarrow \mathbb{C}$, such that $\omega (A^*A)\geq 0 \quad \forall A\in \mathcal{A} $ and $\omega (I)=1$ where $I$ is the identity in $\mathcal{A}$.
\end{definition}
In the cases we are interested in it is possible to focus our attention on a particular class of states, the so called quasi-free states. A quasi-free state can be complitely expressed in terms of its two-point function. For physical reasons, we require this two-point function to be of Hadamard form. The Hadamard condition fixes the singular support of the two-point function and makes possible the regularization procedure for an interacting theory.
\begin{definition}
A Hadamard two-point function is a bidistribution $\omega (x,y)$ whose wavefront set is contained in:
\[
  \{ (x,k_x,y,-k_y)\in T^*(M)^2\setminus \{ 0\} | (x,k_x)\sim (y,k_y),k_x\in V^+_x \}
\]
where $(x,k_x)\sim (y,k_y)$ means that $x$ and $y$ can be joined by a null geodesic and $k_x$ and $k_y$ are cotangent and coparallel to that null geodesic, $M$ is the Minkowski spacetime and $V^+_x$ the future lightcone of $x\in M$.
\end{definition}
In order to give the explicit expression of a state in terms of its two-point function we need some further definitions. We need first to give a notion of dynamics, which allows us to give the definition of a ground state and of a KMS state: In particular, the dynamics of an observable $A\in \mathcal{A}$ is given by a strongly continuous one-parameter group of $*$-isomorphisms $\alpha_t:\mathcal{A}\rightarrow \mathcal{A}$. So:
\begin{definition}
The state $\omega $ is ground (see \cite{Sahl}) if the map $t\mapsto \omega (A\alpha _t(B))$ is such that
\[
  \int _{-\infty }^\infty \widehat{f}(t)\omega (A\alpha _t (B))dt =0
\]
for each $A,B\in \mathcal{A}(M)$, $f\in C_0^\infty (\mathbb{R}^-)$, with $\{ \alpha _t\}_{t\in \mathbb{R}} $ a strongly continuous one-parameter $*$-isomorphism of $\mathcal{A}$.
\end{definition}
\begin{definition}
The state $\omega $ is KMS at inverse temperature $\beta $ if:
\begin{itemize}
\item The functions $t\mapsto \omega (A \alpha _t(B))$ and $t\mapsto \omega (\alpha _t(B)A)$ have an analytic extension to the strip $0<Imz<\beta $ and $-\beta <Imz<0$ respectively;
\item $\omega (A \alpha _t(B))=\omega (\alpha _{t+i\beta }(B)A) \qquad \forall A,B\in \mathcal{A}(M)$
\end{itemize}
\end{definition}

The last important tool to introduce is the time-slice axiom (TSA). We need to build the algebra of observables on the whole spacetime, but the interaction makes the theory divergent. The solution is provided by the TSA: The algebra in a time slice (i.e. a geodesically convex neighborhood of a Cauchy surface of the spacetime) is isomorphic to the one on the whole spacetime. In this way, all the interesting physical objects remain finite.

Let us state it in details. Let $O\subset M$ be a convex open set such that $O\supset \Sigma$ is a Cauchy surface of $M$. Let $\mathcal{A}(M)$ and $\mathcal{A}(O)$ be the algebras of observables over $M$ and $O$ respectively. Suppose that $\mathcal{A} $ is generated by the elements $[f]\in \mathcal{D}(M)/P[D(M)]$. If we introduce a smooth function $\chi ^+$ such that $\chi ^+=1$ in $J^+(O)\setminus{O}$, $\chi ^+=0$ in $J^-(O)\setminus{O}$ and call $E$ the causal propagator of $P$ ($E=E^+-E^-$, $E^\pm $ advanced/retarded fundamental solution of $P$), then $\mathcal{A}(M)$ and $\mathcal{A}(O)$ are $*$-isomorphic via the map:
\[
  f\mapsto P\chi^+ E(f) \qquad f\in C^\infty _0 (M)
\]

All this framework was built for the free theory, but it is possible to extend it to the interacting case perturbatively. In particular (see \cite{FalkThesis}), suppose that the interaction is described by an element $\mathcal{H}_I\in \mathcal{A}$. We can chose as an explicit example the case $\mathcal{H}_I=\lambda \phi ^3$. One can introduce the relative S-matrix (with $T$ time-ordering operator):
\[
  S(\lambda )=\sum_{n=0}^\infty \frac{(-1)^n}{n!}\int _{M^n}d^4x_1...d^4x_n \times T\mathcal{H}_I(x_1)...\mathcal{H}_I(x_n)\lambda (x_1)...\lambda (x_n)
\]
and build the interacting $*$-algebra as the algebra $\mathcal{A}_\lambda (O)$ generated by $S_\lambda (f)=S(\lambda )^{-1}S(\lambda +f)$. It is also possible to express the interacting dynamics $\alpha ^I_t$ in terms of the free one $\alpha _t$, by a co-cycle:
\[
  \alpha _t^I(A)=W_h(t)\alpha _t(A)W_h(t)^{*-1}
\]
with $W_h(t)$ co-cycle defined as a power series of the free dynamics (for details, see \cite{FalkThesis}). Here we introduced a smooth spatial cut-off $h(x)$ equal to $1$ in a compact region of $\Sigma $, which will be sent to $1$ over the whole Cauchy surface later (adiabatic limit).

Introducing the connected correlation functions $\omega _\beta ^C$, the interacting KMS state (smeared with the spatial cut-off $h(x)$) is given by:
\begin{multline*}
		\omega _\beta ^{I,h}(A)=\sum _{n=0}^{\infty} (-1)^n \int _{\beta S_n}du_1...du_n\int_{\Sigma^n}d^3x_1...d^3x_n\times \\
\times h(x_1)...h(x_n)\omega _\beta ^C(A\otimes \mathcal{U}_h(u_1,x_1)\otimes ...\otimes \mathcal{U}_h(u_n,x_n))
\end{multline*}
where:
\begin{itemize}
\item $\beta S_n=\{ (u_1,...,u_n)\in \mathbb{R} |0< u_1< ...< u_n< \beta \}$
\item $\mathcal{U}_h(u,x)=\int dt \dot{\chi }^-(t)\alpha _{iu}([\mathcal{H}_I(x)]_{h\chi })$
\item $\chi^-, \chi \text{ such that }\chi =1-\chi^+-\chi^-$
\end{itemize}

The adiabatic limit corresponds to the limit $h(x)=1$ over $\Sigma \subset O$. As proved in \cite{chil}, the TSA is valid also in the interacting theory, so we are allowd not to care about the divergences in the time direction, considering the $*$-algebra on a time slice $O$ of a Cauchy surface and extending it to the whole spacetime via the time-slice axiom.

Now we can state the most important result of this section. We can express (\cite{FalkThesis}) the expectation value of an observable $A\in \mathcal{A}(O)$ in the ground (KMS) state $\omega _{gr}$ ($\omega _\beta $) in terms of the connected correlation functions
\[
  F^{gr}_{n,G}(u_1,z_1;...;u_n,z_n)= \newline \int dXdY \prod _l G^+_{gr}(x_l-y_l) \Psi (X,Y)
\]
with:
\begin{itemize}
	\item $G^+_{gr}(x)=\frac{1}{2\pi } \int \frac{d^3p }{2\omega _p}e^{-i(\omega _px^0-px)} \qquad (\omega _p=\sqrt{p^2+m^2})$
	\item $\Psi (X,Y)\!=\!\!\prod\limits_{l\in E(G)}\frac{\delta ^2}{\delta \phi _{s(l)}(x_l)\delta \phi_{r(l)}(y_l)}(A_0\!\otimes\! ...\otimes\alpha _{iu_n,z_n}A_n)|_{\phi _i=0}$
	\item $X=(x_1,...,x_n)$, $Y=(y_1,..,.y_n)$
\end{itemize}
Since $F^{gr}_{n,G} \in L^1(\beta S_n\times \Sigma )$, the adiabatic limit $\lim_{h\rightarrow 1}\omega _{gr}^{I,h}(A)$ exists and defines a ground (KMS) state $\omega ^I_{gr}(A)$ ($\omega ^I_\beta (A)$) over $\mathcal{A}(O)$ and then, via the TSA, over $\mathcal{A}(M)$.


\section{Schwarzschild spacetime} \label{schwarz}

In the previous section we saw how things work on flat spacetime. Now we want to switch on the gravity, focusing on the particular case of Schwarzschild spacetime. The main difficulty is to prove the convergence of the two-point function in order to give sense to the expectation values of the observables. For this particular spacetime a work by D. G. Boulware (see \cite{Boul}) gives us the asymptotic behavior of the two-point function of the vacuum (ground) state. Working on this we will be able to build the vacuum state and to generalize it consistently to the case of a system at finite temperature $T$. According to \cite{Boul}, one can expand the solution $\Phi (x)$ of the Klein-Gordon equation on Schwarzschild spacetime:
\[
  (\partial _\mu g^{\mu \nu }\sqrt{-g}\partial _\nu +m^2\sqrt{-g})\Phi (x) =0
\]
in the form:
\[
  \Phi (x)=\sum _{l,m} Y_l^m(\vartheta, \varphi ) \int_{-\infty }^\infty \frac{d\omega }{2\pi }e^{-i\omega t}\phi (r,\omega ;l,m)
\]
Two linearly independent radial solutions $\phi(r,\omega )$ and $\psi(r,\omega )$ can be found, such that their asymptotic behaviour is:
\[
  \phi ^l(r,\omega )\overset{r\to \infty}{\sim} \frac{e^{i(qr+M/q(2\omega ^2-m^2)\log r)}}{qri^{l+1}} \qquad \psi ^l(r,\omega	)\overset{r\to 2M}{\sim} \left(\frac{\mid r-2M\mid }{2M}\right)^{-2i\omega M}
\]
The two-point function of the ground state can be written as:
	\begin{equation*}
	G_{gr}(r,r',\omega )=
	\frac{1}{W[\phi ,\psi ]}\times
	\begin{cases}
	\phi (r,\omega ) \psi (r',\omega ) \quad r'<r\\
	\phi (r',\omega ) \psi (r,\omega ) \quad r<r'
	\end{cases}
	\end{equation*}
where $W[\phi ,\psi ]$ is the Wronskian of $\phi $ and $\psi $. The asymptotic behavior of $G_{gr}$ is determined by $\phi $ (with $a$, $b$, $c$ constants):
	\[
	\phi(r,\omega )\overset{r\to \infty}{\sim} \frac{a}{i^{l+2}}\frac{e^{-br}}{r}r^{-c} \text{  for } \omega ^2<m^2; \quad  \phi(r,\omega )\overset{r\to \infty}{\sim} \frac{e^{i(qr+a\log r)}}{i^{l+1}qr} \text{  for } \omega ^2>m
	\]
As a consequence, also on Schwarzschild spacetime we have $F^{gr}_{n,G}\in L^1(\beta S_n\times \Sigma )$. Moreover, we still have a consistent definition of ground state with respect to $\alpha ^I_t$ and the TSA (to build $\mathcal{A}(M)$ from $\mathcal{A}(O)$). As a consequence, analogously to the flat case, the adiabatic limit $\lim_{h \to 1}\omega _{gr}^{I,h}(A)=\omega _{gr}^I(A)$ exists and defines a ground state on the static region of Schwarzschild spacetime.

For a KMS state at inverse temperature $\beta $ the calculations are basically the same. The only difference is that a Bose factor appears in the two-point function, which takes the form:
\[
  G_\beta (x,x')=i\int _0 ^\infty d\omega  \frac{e^{-i\omega (t-t')}}{1-e^{-\beta \omega }} G_{gr}(r,r',\omega )
\]
This does not affect the converging properties of the state, so the adiabatic limit can be performed also in this case.

The last remark concerns some possible generalizations to larger classes of spacetimes. Since we are looking for asymptotical convergence, we can suppose that the general procedure showed here would be still valid for all the stationary asymptotically well-behaved spacetimes, and in particular we would start to work on stationary asymptotically Minkowskian (SAM) spacetimes very soon. A further possible generalization is to higher spin fields and to the massless case.


\section{Conclusion}

One of the problems in extending QFT to curved spacetime is to keep the asymptotical behaviour of the theory under control, avoiding divergences due to the particular geometry of the spacetime. Here we showed how this can be done for the Schwarzschild spacetime taking advantage of the algebraic approach, already used by other autors for a new approach to QFT also on the flat spacetime. We also showed some possible extensions to a larger class of spacetimes and fields.

\end{document}